\documentclass{ws-p8-50x6-00}
\usepackage{epsfig,here}
\def\SDG{\hbox{\tiny SDG}}
\begin{document}

\title{Resummation and power corrections\\ by Dressed Gluon Exponentiation}

\author{Einan Gardi} 

\address{TH Division, CERN, CH-1211 Geneva 23, Switzerland\\
E-mail: {\tt Einan.Gardi@cern.ch}}  

\author{Johan Rathsman}
\address{Department of Radiation Sciences, Box 535, S-751 21 Uppsala, Sweden\\
E-mail: {\tt rathsman@tsl.uu.se}}

\maketitle

\abstracts{Event-shape distributions in $e^+e^-$ annihilation offer a unique laboratory for understanding perturbative and non-perturbative aspects of QCD.
Dressed Gluon Exponentiation is a resummation method designed to evaluate differential cross sections close to a kinematic threshold and provide the basis for parametrization of power corrections. The method 
and its application in the case of the thrust and the heavy jet mass distributions in the two-jet region are briefly presented.}

Event-shape distributions are an effective way to describe the 
final state in hard processes. In the case of $e^+e^-$ annihilation, event shapes offer a unique laboratory for understanding perturbative and non-perturbative aspects of the theory. Being infrared- and collinear-safe quantities, event-shape distributions can be calculated perturbatively. 
Nevertheless, these observables typically have large perturbative and non-perturbative corrections, which are related to fundamental properties of soft gluon radiation and hadronization.   

Particularly interesting physics appears in the two-jet region, where the cross-section peaks. Then $1\,-$ thrust, $t$, is small, and equals, to a good approximation, the sum of the two (light and heavy) jet masses $t\simeq \rho_L+\rho_H$. Any hard emission, which makes the variables $t$ and $\rho_H$ large, is 
suppressed by $\alpha_s$ at the hard scale~$Q$.
In the two-jet region the cross section is dictated by soft and collinear gluon radiation. This has two immediate implications. The first is that, in spite of the large centre-of-mass energy $Q$, the relevant scale in the two-jet region is low, $\sim t Q$. The second is that, because of the singularity of the matrix element, multiple emission must be taken into account.
Consequently, in the two-jet region fixed-order calculations are insufficient.
Two techniques were developed to confront such difficulties:~(1)~renormalons$^{2-9}$,
 which were mainly used to identify power corrections through 
the sensitivity of the perturbative calculation to low momenta. 
This sensitivity is also reflected in the divergence of the 
perturbative expansion at large orders; and~(2)~Sukakov resummation\cite{Catani:1993ua}, which deals with the summation of logarithmically enhanced contributions that dominate the perturbative coefficients close to a kinematic threshold. 

The two dominant characteristics of the two-jet region, namely low momenta
and multiple emission are addressed by Dressed Gluon Exponentiation\cite{Thrust_distribution}~(DGE). The calculation of the distribution can be seen as a two-step procedure, where the first step is a typical renormalon calculation, namely the evaluation 
of a {\em single} dressed gluon (SDG) cross-section, and the second is the exponentiation. The SDG calculation is similar to a leading order calculation, where a gluon is emitted from the primary quark--antiquark pair, with the essential difference that the gluon can be off-shell and the actual gluon virtuality is used as the scale of the running coupling. It is in this 
that DGE differs from previous calculations, such as the resummation with next-to-leading logarithmic (NLL) accuracy\cite{Catani:1993ua}. The exponentiation 
itself relies on the factorization property of soft and collinear radiation, which holds in the case of an off-shell gluon\cite{DGE}. DGE makes the same approximation as NLL resummation that gluons are emitted independently.

The DGE expression for the single jet mass distribution is given by an inverse Laplace integral:
\begin{eqnarray}
\frac{1}{\sigma}\frac{d\sigma}{d\rho}(\rho,Q^2)=\int_{c}
\frac{d\nu}{2\pi i} \exp \left\{\rho\nu+ \ln J_{\nu}(Q^2)\right\},
\end{eqnarray}
where $c$ is an integration contour parallel to the imaginary axis and
\begin{eqnarray}
\label{lnJ}
\ln J_{\nu}(Q^2)&=& \nonumber
\int d\rho \left. \frac{1}{\sigma}\frac{d\sigma}{d\rho} (\rho,Q^2)\right\vert_{\SDG}
\left(  e^{-\nu \rho}-1 \right)\\
&=&\frac{C_F}{2\beta_0}
\int_0^{\infty} d{z}\,B_{\nu}({z}) \exp\left(-{{z} \ln Q^2/\bar{\Lambda}^2}\right)\,
 \frac{\sin\pi{z}}{\pi{z}}. 
\end{eqnarray}
Here $\bar{\Lambda}$ corresponds to the ``gluon bremsstrahlung'' effective charge and the Borel function is given by
\begin{eqnarray*}
\,B_{\nu}({z})
&=&\frac{2}{{z}}\left[e^{2{z}\ln\nu}\Gamma(-2{z})
+\frac{1}{2{z}}\right]-\left(\frac{2}{{z}}+\frac{1}{1-{z}}
+\frac{1}{2-{z}}\right)\left[e^{{z}\ln \nu}\Gamma(-{z})
+\frac{1}{{z}}\right].
\label{Borel_nu}
\end{eqnarray*}
Using this distribution, and the assumption that the hemisphere masses are independent, both the thrust and the $\rho_H$ distributions are readily obtained,
\begin{eqnarray}
\label{t}
\frac{1}{\sigma}\frac{d\sigma}{dt}(t,Q^2)
&=&\frac{d}{dt}\int_c\frac{d\nu}{2\pi i
\nu}\,\exp\left\{\nu t+  2  \ln J_\nu(Q^2)\right\}\\
\label{rho_H}
\frac{1}{\sigma}\frac{d\sigma}{d\rho_H}(\rho_H,Q^2)
&=&\frac{d}{d\rho_H}
\left[\int_c\frac{d\nu}{2\pi i\nu}
\,\exp\left\{\nu\rho_H+ \ln J_\nu(Q^2)\right\}\right]^{ 2 }.
\end{eqnarray}
Since only logarithmically enhanced terms are resummed by the DGE formulae, matching\cite{Catani:1993ua} with fixed-order results is necessary. The currently available accuracy of such calculations is next-to-leading order (NLO).

At the perturbative level, DGE differs from the standard NLL resummation by a class of formally sub-leading logarithms, which are factorially enhanced. This enhancement is a direct consequence of the infrared renormalons: the numerical factors emerge from the integration over the QCD coupling at low momenta. 
These corrections cannot be neglected: at $Q= M_{\rm Z}$ they amount to $\sim 20\%$ of the exponent $\ln J_{\nu}(Q^2)$.  
The resummation of running-coupling effects in the exponent makes it renormalization-scale invariant, thus avoiding a major source of uncertainty.

The presence of power-suppressed ambiguities in (\ref{lnJ}) indicates 
specific non-perturbative corrections. The appearance of these infrared renormalons in the exponent implies that the associated 
power corrections exponentiate\cite{Thrust_distribution}. Moreover, assuming the dominance of these corrections, the dependence of the ambiguity on $\nu\sim 1/\rho$ is indicative of the functional form of the power corrections\cite{Thrust_distribution}. The most important Borel 
singularities in $\ln J_{\nu}(Q^2)$ appear at half integer values of 
$z$ and are related to soft gluon emission at large angles.
The first singularity, at $z=1/2$, corresponds to a shift of the distribution\cite{Dokshitzer:1997ew}, while the main effect of higher singularities can be recast\cite{Korchemsky:1999kt,Korchemsky:2000kp,Belitsky:2001ij} in a  convolution of the perturbative distribution with a shape function (SF)
that depends on a single variable $\rho Q$. Finally, the absence of Borel singularities of this type at 
integer~$z$ values suggests\cite{Thrust_distribution} that the even central moments of the SF are suppressed, while the relevant non-perturbative parameters are the odd ones. This explains, in particular, why a shift\cite{Dokshitzer:1997ew} can be a good approximation to the SF in a wide range of thrust (or $\rho_H$) values. 

In the approximation considered, both the thrust (\ref{t}) and the $\rho_H$ (\ref{rho_H}) distributions are expressed in terms of the single jet mass distribution~$ J_{\nu}(Q^2)$. 
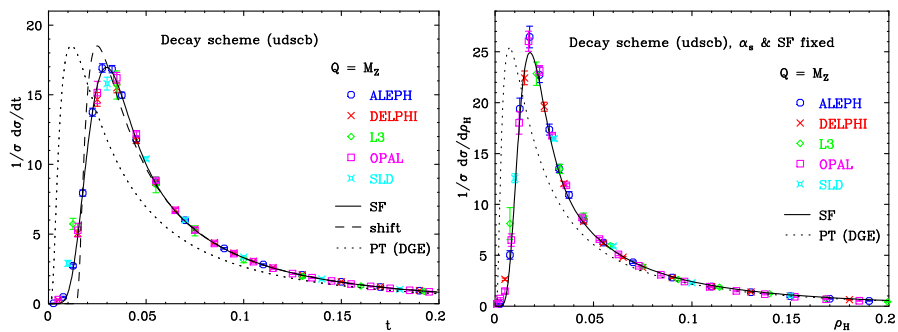
\begin{figure}[H]
  \begin{center} 
\epsfig{width=4.29truecm,angle=90,file=thrust_kcorr_mz.ps}
\epsfig{width=4.29truecm,angle=90,file=rhoh_kcorr_mz_allfix.ps}
\end{center}
\end{figure}
\noindent 
Therefore, a single SF incorporates the non-perturbative 
corrections in both~cases.
This implies that fixing the unknown parameters ($\alpha_s$ and the SF)
by a fit to the thrust distribution, the  
non-perturbative $\rho_H$ distribution can be simply calculated and compared with experimental data. The results of this exercise are presented in the above figures\cite{Jet_mass}.
The figure on the left shows the thrust distribution data at $M_{\rm Z}$
compared with the perturbative DGE calculation~(\ref{t}) (dotted line) as well as a shift (dashed line) and a SF (full line) fits. The fits, which take into account data in the range 
$14$ -- $189$ GeV, yield 
$\alpha_{\overline{\mbox{\tiny MS}}}(M_{\rm Z}) = 0.1090 \pm 0.0004$, the first two central moments of the SF being
$\lambda_1\bar{\Lambda} = 0.605\pm 0.013 \; \; {\rm GeV}$ and
$\lambda_2\bar{\Lambda}^2 = 0.002\pm 0.024 \; \; {\rm GeV}^2$.
The errors quoted are experimental. The figure on the right shows the 
\hbox{{\em prediction}} for the $\rho_H$ distribution, which uses as input the parameters of the thrust best fit, compared with data. 
The overall agreement is good.
This result supports our approximations and, in particular, it supports 
our assumption that the correlation between the hemispheres is insignificant 
in the peak region.
Nevertheless, two major caveats must be pointed out. 
The first is that the data used do not correspond to the standard hadronic level, but rather to a measurement 
where all hadrons are assumed to decay into massless particles.  
The ``decay scheme''
is an effective way to decrease the effect of hadron masses on the measured distribution\cite{Salam:2001bd,Jet_mass}. 
The second caveat is that, in the $\rho_H$ case, an independent fit where $\alpha_s$ is free yields a significantly lower best-fit value for~$\alpha_s$ 
and a correspondingly higher value for $\lambda_1$. 
This is shown in the following figure\cite{Jet_mass}, which summarizes the best shift-based fits for the thrust and the $\rho_H$~distributions. 
Results of independent fits, where both~$\alpha_s$ and the shift~$\lambda_1$ are fitted, are shown as one $\sigma$ ellipses, while fits where~$\alpha_s$ is 
fixed are shown as crosses. 
The $\rho_H$ distribution fits are sensitive to the fitting~range.~Agreement
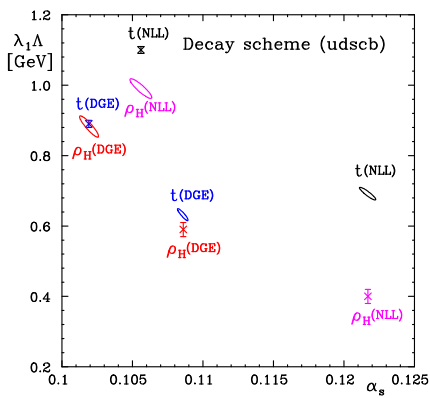
\begin{figure}[H]
  \begin{center}
    \epsfig{width=5.16truecm,angle=90,file=correlation.ps}
  \end{center}
\end{figure}
\noindent 
with a SF based fit, which is much more stable, was used to optimize the fitting range. The sensitivity to the fitting range provides a strong indication that the perturbative approximation to the $\rho_H$ distribution is not good enough. This also explains the discrepancy 
in~$\alpha_s$ with respect to the thrust.   
A~detailed analysis shows\cite{Jet_mass} that the missing perturbative corrections are primarily associated with non-logarithmic terms at NNLO, which would tend to decrease the cross section at large~$\rho_H$. 
Even in the absence of a NNLO result, a~quantitative analysis of power corrections 
{\em in the peak region}, based on the DGE resummation formula and the available NLO, is possible. Indeed, as the figure shows, when~$\alpha_s$ is fixed according to the thrust fit, the $\rho_H$ best-fit value for~$\lambda_1$ agrees with that of the thrust.
The agreement is due to the improved approximation achieved by DGE: the discrepancy in a NLL based fit is large.

We conclude that incorporating the resummation of running-coupling effects in the Sudakov exponent is essential for a quantitative analysis of the distribution in the two-jet region. Since power corrections are important, 
resummation with a fixed logarithmic accuracy is insufficient. 
DGE is therefore indispensable.


\begin{thebibliography}{99}

\bibitem{Catani:1993ua}
S.~Catani, L.~Trentadue, G.~Turnock and B.~R.~Webber,
{\em Nucl. Phys.}  {\bf B407} (1993) 3.

\bibitem{Dokshitzer:1995zt}
Y.~L.~Dokshitzer and B.~R.~Webber,
{\em Phys. Lett.}  {\bf B352} (1995) 451.

\bibitem{Akhoury:1995sp}
R.~Akhoury and V.~I.~Zakharov,
{\em Phys. Lett.}  {\bf B357} (1995) 646.

\bibitem{DMW}
Y.~L.~Dokshitzer, G.~Marchesini and B.~R.~Webber,
{\em Nucl. Phys.}  {\bf B469} (1996) 93.

\bibitem{Average_thrust}
E.~Gardi and G.~Grunberg,
{\em JHEP} {\bf 9911} (1999) 016.

\bibitem{Gardi:2000yh}
E.~Gardi,
{\em JHEP} {\bf 0004} (2000) 030.

\bibitem{Dokshitzer:1997ew}
Y.~L.~Dokshitzer and B.~R.~Webber,
{\em Phys. Lett.}  {\bf B404} (1997) 321.

\bibitem{Korchemsky:1999kt}
G.~P.~Korchemsky and G.~Sterman,
{\em Nucl. Phys.}  {\bf B555} (1999) 335.

\bibitem{Korchemsky:2000kp}
G.~P.~Korchemsky and S.~Tafat,
{\em JHEP} {\bf 0010} (2000) 010.

\bibitem{Belitsky:2001ij}
A.~V.~Belitsky, G.~P.~Korchemsky and G.~Sterman,
{\em Phys. Lett.}  {\bf B515} (2001) 297.

\bibitem{DGE}
E.~Gardi,
``Dressed gluon exponentiation'', to appear in {\em Nucl. Phys.} {\bf B}, [hep-ph/0108222].

\bibitem{Thrust_distribution}
E.~Gardi and J.~Rathsman,
{\em Nucl. Phys.}  {\bf B609} (2001) 123.

\bibitem{Jet_mass}
E.~Gardi and J.~Rathsman,
``The thrust and the heavy jet mass distributions in the two-jet region'',
to be published.

\bibitem{Salam:2001bd}
G.~P.~Salam and D.~Wicke,
{\em JHEP} {\bf 0105} (2001) 061.


\end{thebibliography}
\end{document}